\documentclass[preprint]{iucr}              

     \journalcode{B}              

\begin{document}                  



\title{Charge-density-wave state induced by structural distortion in heavy-fermion compounds Ce$_3$M$_4$Sn$_{13}$ (M=Co, Ru, Rh).}


\cauthor[a]{Lech}{Kalinowski}{lkalinowski@us.edu.pl}{address if different from \aff}
\author[a]{Jerzy}{Goraus}
\author[a]{Andrzej}{\'Slebarski}

\aff[a]{Institute of Physics, University of Silesia, Uniwersytecka 4, 40-007 Katowice, \country{Poland}}









\maketitle                        

\begin{abstract}
In previous reports a structural transition from a cubic phase of Yb$_3$Rh$_4$Sn$_{13}$ - type to the superlattice variant has been reported at $\sim160$ K for a series of skutterudite-related Ce$_3$M$_4$Sn$_{13}$ compounds, where M=Co, Ru or Rh. We have simulated the low-temperature XRD diffraction patterns of the distorted unit cell using written for that purpose DISTorX program. The method proposed here for x-ray diffraction analysis obtains the XRD patterns from the atomic positions and allows to investigate crystal structure without imposed symmetry operations. We have indicated crystallographic plane where distortion occurs and explained possible origin of CDW in these materials. We have also shown that distortion caused by the charge density wave leads to significantly changes of the intensity of diffraction lines.
\end{abstract}

\section{Introduction}

Filled skutterdites with a chemical formula R$_3$M$_4$Sn$_{13}$, where R is rare earth atom, M is Co, Ru or Rh, have been extensively investigated due to their diverse heavy fermion properties, also including interesting interplay of superconductivity and magnetic order and thermoelectric properties closely related to the cage-type structure (Aoki et al. 2005, Maple 2005, Sales 2003). The Sn1 ions  form a bcc structure in the Yb$_3$Rh$_4$Sn$_{13}$ - type compounds. Each Sn1 atom is surrounded by 12 pnictogen Sn2 ions forming slightly distorted icosahedron, and eight transition elements forming a cube. Each atom M is located inside M(Sn2)$_6$ trigonal prism, while R atom centers R(Sn2)$_{12}$ cuboctahedra. The recent density functional calculations (Gam\.za et al. 2008, Zhong et al. 2009) showed high charge density accumulation between metal M and Sn2 atoms, which implies a strong covalent bonding interaction. It has been shown (\'Slebarski et al. 2012) that even a small change of local symmetry, e. g., of the M(Sn2)$_6$ and R(Sn2)$_{12}$ cages, generated by small deformation of the Sn$_{12}$ icosachedra can lead to variation in the charge density of metal M and R ions. The slight structural deformation of the R(Sn2)$_{12}$ cage also changes the symmetry of crystal electric field (CEF), which is evidently observed in the specific heat and magnetic susceptibility data (\'Slebarski et. al. 2014a). Generally, the strong covalent bonding in R-based skutterudite-related compounds can be a key mechanism in the variety of unusual properties (Goto et. al. 2005; Thomas et. al. 2006). In particulare, the heavy fermion properties for R=Ce (\'Slebarski et al. 2012), superconductivity with novel structural quantum critical point (QCP) for R=La (Klintberg et. al. 2012, \'Slebarski et al. 2014c, Chenng et al. 2016), and strong electronic correlation have attracted considerable attention, due to the electronic structure and cage-type crystal structure.  A detailed investigations and empirical analysis suggest that structural distortion, which is observed at $T_D \sim 160$ K in most of these compounds (Lue et al. 2012, Cheung et al. 2016), has tremendous influence on the electronic structure of the system (\'Slebarski \& Goraus 2013, \'Slebarski et al. 2014b). Moreover, strong electron correlations induce anomalous behavior of the low temperature specific heat with characteristic C(T) dependences. For cubic CEF the ground state of Ce - 4f level should be properly described by two degenerate energy states, the lowest doublet and the highest quartet, or inversely. It was shown experimentally that the distortion of the Ce cubic cages induces the change of CEF effect, and the best description of the ground state properties can be obtained for the CEF with tetragonal symmetry (\'Slebarski et al. 2014a). The La counterparts are BCS superconductors (\'Slebarski et al. 2014c) with hypothetical structural quantum critical point at high external pressure of about 20 GPa. Strongly correlated electronic systems (SCES), e. g., Ce$_3$Co$_4$Sn$_{13}$  exhibit the presence of charge density wave, with accompanying structural distortion (Lue et al. 2012). Here we present the low temperature x-ray diffraction patterns and the simulated ones, considering structural distortion caused by charge density wave. This structural phase transition has also been observed recently at $T_D \sim 160$ K for Ce$_3$Ru$_4$Sn$_{13}$ and Ce$_3$Rh$_4$Sn$_{13}$ compounds (\'Slebarski \& Goraus 2013), similar to CDW-type ordering suggested by Song et al. 2003. Here, we investigate the effect of CDW structural change on diffraction patterns of the series of Ce$_3$M$_4$Sn$_{13}$ compounds. This procedure gives us a chance to compare and verify a model structure with the experimental data. We also determined the crystallographic planes responsible for the structural distortion in R$_3$M$_4$Sn$_{13}$, assuming that the amplitude of CDW could be a reason of structural disorder. The observed structural disorder seems to be essential to understand competition between superconductivity and CDW in family of La-based R$_3$M$_4$Sn$_{13}$ compounds (Balseiro \& Falicov 1979), as well as the nature of structural quantum phase transitions at $T_D \rightarrow 0$.

\section{Experimental Details}

Polycrystalline samples Ce$_3$M$_4$Sn$_{13}$, M=Co, Ru and Rh, have been prepared by arc melting of the pure elements on water-cooled copper hearth in a high-purity argon atmosphere with an Al getter. Samples were remelted several times to favor homogeneity and annealed at 870$^oC$ for 14 days. All samples were carefully examined by x-ray-diffraction analysis at room temperature and appeared as single phase with a cubic structure (space group $Pm\bar3n$). Low temperature x-ray diffraction was performed using the high-resolution diffractometer Empyrean PANalitical B. V. with low temperature detector TTK 450 Anton-Paar. Refinements of Bragg peaks in the scattering angle range $10^o \le 2 \theta \le 90^o$ yield the lattice patameter $a$ for each compound with the accuracy higher then $2 \times 10^{-4}$\AA. Stoichiometry of the samples were checked by x-ray fluorescence spectrometer ZSX Primus II Rigaku (EDXRF) equipped with x-ray tube with rhodium anode. X-ray photoelectron spectra (XPS) were obtained with monochromatized Al $K_{\alpha}$ radiation at room temperature using a PHI 5700 ESCA spectrometer. XPS survey analysis and XRF results, confirm small deviation of nominal composition at the level $\sim 1\%$. 

\section{CDW simulation method}

Optimization of input structure was based on Hellmann-Feynman theorem (Wu \& Van Voorhis 2005, Hine et al. 2011), applied with DFT calculation method. The structures were visualized using VESTA program (Momma \& Izumi 2008). XRD Ritveld analysis was accomplished using commercial licenced Xpert High Score Plus program. 
XRD simulations are based on Bragg and Laue's and Fourier's theoretical basis of solid state physics. DISTorX computer program used to simulate XRD patterns was written in Python 2.7 programming language (Python Software Foudation 2001) using "pylab", "numpy" and "cmath" libraries. Visualization module (Scherer et al. 2000) of simulation program was also helpful in identification of atomic positions, and animated model of CDW distortion. Simulated x-ray patterns are displayed using "matplotlib" library (Hunter 2007) with dynamic plot module, where XRD patterns  are dynamically changed with changes of CDW amplitude as animation. All modules are connected using "Tkinter" graphical user interface (GUI) to manage simulation results and to simplify the entry method of initial parameters. Our program DISTorX is published in the on-line repository (Kalinowski 2016). Line profile was introduce as Pseudo-Voigdt distribution function. XRD patterns simulated with DISTorX were compared with other well respected software programs namely: Full Prof package (Rodriguez-Carvajal 1993), Powder Cell (Kraus \& Nolze 1996) and Mercury (Macrae et al. 2008) programs.
Experimental diffraction patterns were refined by Ritveld method for all samples at $300$ K and $85$ K, the results are presented in Table 1. Refined lattice parameters are used in CDW distortion simulation. The model implies change in positions of atoms' across $c$ crystal axis. Changes in environment of the R atom lead to strong reconstruction of the Fermi level and the electronic structure, as was shown in previous reports (\'Slebarski \& Goraus 2013, \'Slebarski et al. 2014b). We assumed presence of the charge accumulation in the 
(0 0 1) planes near Ce atoms. In order to achieve that charge accumulation one has to consider a primitive unit cell with doubled lattice parameter c. The considered distortion also explains the observed superlattice peaks in the XRD patterns. In that case, two Ce atoms are located directly in the middle of the modified unit cell. We modeled the distortion with
\begin{equation}
Z_N(i) = Z(i) + A\cos(2\pi Z(i)),
\end{equation}
where, $Z_N(i)$ represents new $z$ Cartesian coordinate of i$^{th}$ atom, $Z(i)$ is an old $z$ coordinate of an atom and $A$ is an amplitude of the charge density wave. Figure 1 shows the structural distortion for the modified unit cell. In this model the Ce-atom positions are stable, while M and Sn atoms have a degree of freedom and can be shifted. Our program generates XRD patterns from direct Cartesian $(x, y, z)$ positions of atoms, treating input structures as $P1$ symmetry. This allows us to study structures without imposed symmetry operations and thus, allows to generate diffraction patterns with arbitrarily determined distortion. DISTorX computes all possible diffraction lines, both, caused by distortion and coming from the parent $Pm\bar3n$ structure of the base compound. Atomic form factors were applied as a function of tabulated parameters $a_n$ of x-ray scattering ability of elements (Brown et al. 2006) 
\begin{equation}
f = \sum_{n=1}^{m} a_{(2n)} \exp [-a_{(2n+1)}y^2],
\end{equation}
where $y = k/4\pi$, and $k$ represent the length of reciprocal lattice vector. Structure factors were calculated to obtain the intensities of the diffraction lines and 
\begin{equation}
I = |F|^2 * \frac{1+cos^22\theta}{sin^2\theta cos\theta},
\end{equation}
where the structural F-factor is multiplied by the Lorentz polarization factor.

The simulated diffraction patters were obtained for different Ce$_3$M$_{4}$Sn$_{13}$ compounds independently from: Full-Prof, Powder-Cell and Mercury to compare, if each of them gives similar result and compared to the pattern generated by DISTorX. All the simulated patterns showed similar and good enough match to the experimental data.

DISTorX code was written in a procedural programing manner. This allows to use parts of code separetly and provide to run only XRD simulation functions using web based python interpreter like Jupyter (Jupyter 2016) or Sage (Stein \& Joyner 2005). The same programing manner is applied to other parts of a program like as simulation of CDW or visualization part. This type of programing method gives us thoroughly system independent code. Program DISTorX, was tested either on Windows operation systems (Windows 7, 8, 8.1, 10) and Linux distributions (Gento, Linux Mint). Figure 2 shows graphical user interface (GUI) of written program. Left side includes access to main functions, and simulation-experiment comparison module. The right side of the window shows parameters input and buttons providing to functions like normalization of experimental-simulation intensities. Figure 3 shows schematic operation of the program.

\section{Results and Discussion}

The simulation obtained on the optimized and doubled unit cell with applied CDW distortion gives an additional $2\theta \sim 24.3^o$ reflection, which is not present in other simulated patterns generated from the base atomic Wyckoff positions with applied symmetry operations. This superlattice diffraction line well interprets the XRD pattern obtained at the temperatures lower than $T_D$ in the $2\theta < 40^o$ diffraction angles. That could confirms the presence of superlattice variant with double $c$ lattice parameter visible in experiment, and confirm that the structural phase transition might be caused by charge density wave. Now we discuss the distortion lattice plane in Ce$_3$Rh$_{4}$Sn$_{13}$, visible in low temperature diffractiom pattern and determined on the base of DISTorX programe.
Figure 4 compares the XRD patterns measured for Ce$_3$Co$_{4}$Sn$_{13}$, Ce$_3$Ru$_{4}$Sn$_{13}$, and Ce$_3$Rh$_{4}$Sn$_{13}$, respectively at $T=300$ K and $T=85$ K. At $85$ K the additional peak is clearly visible in all diffraction patterns at $2\theta \sim 24.2^o$ . Moreover, the change of its intensity is strongly M dependent, and has a maximum value for Rh. In case of Ce$_3$Rh$_{4}$Sn$_{13}$ the XRD pattern at $85$ K exhibits also another diffraction line at $2\theta \sim 43.6^o$. All these diffraction patterns are well compared with the simulation of the CDW distorted structure with the amplitude A changed between $0.001$ \AA \ and $1$ \AA. Experimental and simulated intensities are normalized to $1$. The observed change of the distortion amplitude correlates with the change of the intensities, of the diffraction lines, moreover when A value larger then $1$ \AA \ the $2\theta \sim 24.2^o$ diffraction line disappears.

In Figure 4 the black pattern represents structure with the amplitude $A=0$ (i. e., without distortion), whereas cyan pattern represents distorted structure with $A=0.05$ \AA. We noticed that CDW structural distortion generates a new peak located in the XRD diffraction pattern at about $2\theta = 24.2^o$, which is not expected by Ritveld analysis for the base unit cell. For  the amplitude $A=0.05$ \AA \ we obtained the best agreement between the XRD pattern and the simulated one. In commercial Xpert High Score Plus program the Ritveld analysis of the experimental pattern implies a weak diffraction lines at $2\theta = 22.7^o$ (1 1 2), $2\theta = 26.3^o$ (0 2 2) barely seen experimentally. 

At $2\theta = 24.2^o$ we designates equivalent lattice planes (0 1 5) (1 0 5) (2 1 3) shown in figure 5 as blue and (1 2 3) plane (yellow). Plane (1 2 3) is most likely related to structural phase transition. This plane is located between (1 1 2) and (0 2 2) lattice planes, as is shown in figure 5. It intersects icosahedron cages between Sn1-Sn2 atoms, moreover, M atoms lie in this plane. Previous reports (\'Slebarski et. al. 2014b) clearly indicate charge accumulation inside the cages, and charge deficit directly observed on transition metal M. Both effects lead to deformation of the cages. An accompanying subtle structural distortion changes the symmetry of the crystal electric field (Ślebarski et. al. 2014a). The CEF symmetry is lower than cubic which is well documented by the specific heat and magnetic susceptibility data.

\section{Conclusions}

The temperature dependent x-ray diffraction patterns for Ce$_3$M$_4$Sn$_{13}$ documented the presence of structural phase transition at $\sim 160$ K from a cubic $Pm\bar3n$ structure to its superlattice variant. We succesfully simulated x-ray diffraction patterns either above or below $T_D$ using DISTorX. We demonstrated that the distortion caused by the charge density accumulation generate the additional diffraction peaks in XRD patterns at the temperature region $T \le T_D$. We have also shown which planes are more likely to be distorted.  Our previous DFT results (\'Slebarski et. al. 2015a; \'Slebarski et. al. 2015b) indicated the impact of the f-electron and d-electron correlations on the charge distribution in the unit cell of Ce$_3$M$_4$Sn$_{13}$. For M = Co a significant loss of spatial charge on metal M is documented, which is weaker for Rh, in contrast to a positive charge accumulation observed for Ru. Moreover, for M=Co and M=Rh, the Sn2 cages accumulate inside most of the charge between Sn1 and Sn2 atoms. In result the charge accumulation leads to strong covalent bonding. This bonding between M and Sn1 leads to strong distortion in Ce$_3$M$_4$Sn$_{13}$, well manifested in the XRD patterns. Both, the charge accumulation in the cages and charge deficit on the transition metal M give rise to a charge density wave, that is a reason of the structural distorion. The direction of charge density wave propagation is not associated with any privileged crystallographic axis (a, b, c) and may arise along the lattice planes (1 2 3), where the effect of charge accumulation on the shift of metal M is the most significant.

We also would like to draw an attention on universality of presented CDW simulation method. We present the tool which can be use as whole program using GUI, or as a part of code in web based interpreter. There are no limits for input structures, and users with lack of programing knowledge can model and test many of possible distortions on various classes of materials.
 
\ack{Acknowledgments}

We thank the National Science Center (NCN) for financial support on the basis of Decision No. DEC-2012/07/B/ST3/03027 and also for financial support on the basis of agreement No. UMO-2014/15/N/ST3/03799.

\begin{table}
\centering
    \begin{tabular}{lclclclcl}
    \hline
Compound~ &$T$ [K]  ~ & $a$ [\AA] ~ & $2a$ [\AA] ~ & $R_{Bragg}$ \\ \hline
    
Ce$_3$Co$_{4}$Sn$_{13}$~ &$300$ &$9.5949$& 19.1898 & 13.5\\ 
    
 & $85$ & 9.5677 & 19.1355 & 14.4 \\
 
Ce$_3$Ru$_{4}$Sn$_{13}$~ & $300$ & 9.7249 & 19.4498 & 6.2 \\ 
   
 & $85$ & 9.7009 & 19.4018 & 6.0 \\
    
Ce$_3$Rh$_{4}$Sn$_{13}$~ & $300$ &9.5958 & 19.1917 & 11.9 \\ 
    
 & $85$ & 9.5679 & 19.1359 & 12.8 \\ \hline
\\
    \end{tabular}

\textbf{Table 1.} The lattice parameters of Ce$_3$Co$_{4}$Sn$_{13}$, Ce$_3$Ru$_{4}$Sn$_{13}$ and Ce$_3$Rh$_{4}$Sn$_{13}$ respective and the obtained $R_{Bragg}$ factors.

\end{table}

Figure 1. Visualization of doubled unit cell of Ce$_3$M$_4$Sn$_{13}$ (M=Co, Ru, Rh) after optimization with selected symmetry axis. Arrows indicate the direction of deformation caused by CDW. Green balls represents Ce atoms, red corresponds to M metal, and gray represents Sn1 and Sn2 atoms.


Figure 2. General user interface of DISTorX program.


Figure 3. DISTorX working scheme.


Figure 4. Cubic structure $Pm\bar3n$ with the doubled lattice parameter $a=9.7$ \AA \ and applyed CDW distortion visualized in DISTorX program.


Figure 5. XRD experimental diffraction patterns for Ce$_3$Co$_{4}$Sn$_{13}$, Ce$_3$Ru$_{4}$Sn$_{13}$ and Ce$_3$Rh$_{4}$Sn$_{13}$ at $300$ K and $85$ K. Insets show detailed areas where the superlattice peaks arise.


Figure 6. Experimental x-ray diffraction patterns for Ce$_3$Rh$_{4}$Sn$_{13}$ respectively at $300$ K and $85$ K, with Rietveld refinement and difference.


Figure 7. Experimental diffraction pattern for Ce$_3$Rh$_{4}$Sn$_{13}$ at $300$ K and $85$ K, compared with simulation of CDW distortion. Insets display region where superlatice peaks appear.




Figure 8, 9. Visualization of Ce$_3$M$_4$Sn$_{13}$ (M=Co, Ru, Rh) after optimization with marked lattice planes (1 1 2), (0 2 2) green color, (0 1 5), (1 0 5), (2 1 3) - blue color, and (1 2 3) - yellow color. A colorful balls represent atoms described in figure 1.


Figure 10, 11. Visualization of  Ce$_3$M$_4$Sn$_{13}$ (M=Co, Ru, Rh) with marked lattice plane (1 2 3). A colorful balls represent atoms identified in figure 1.


\end{document}